\documentclass[conference, 9pt]{IEEEtran}
\IEEEoverridecommandlockouts

\usepackage{cite}
\usepackage{blindtext, graphicx}
\usepackage[cmex10]{amsmath}
\usepackage{amsfonts,amsthm,amssymb}
\usepackage[font=scriptsize,caption=false,labelsep=space]{subfig}
\usepackage{graphicx,float,wrapfig}
\usepackage{graphicx}
\usepackage{graphics,color}
\usepackage{mathtools}
\usepackage{nicefrac}
\usepackage{bm}
\usepackage{txfonts}

\def\IEEEbibitemsep{0pt plus .5pt}
\usepackage{balance}

\usepackage{color}

\usepackage{wrapfig} 
\usepackage{pgfplots}
\usepackage{upquote}
\usepackage[none]{hyphenat}

\begin{document}
%
\title{Cognitive Sub-Nyquist Hardware Prototype of a Collocated MIMO Radar\vspace{-24pt}}

\author{\IEEEauthorblockN{Kumar Vijay Mishra, Eli Shoshan, Moshe Namer, Maxim Meltsin, David Cohen, \\Ron Madmoni, Shahar Dror, Robert Ifraimov, and Yonina C. Eldar}\thanks{The authors are with the Andrew and Erna Viterbi Faculty of Electrical and Computer Engineering, Technion - Israel Institute of Technology, Haifa, Israel, e-mail: \{mishra@ee, elis@ee, namer@ee, maxim.meltsin@ee, davidco@tx, mron@t2, shahard@t2, rifraimov@ef, yonina@ee\}.technion.ac.il}\thanks{This project has received funding from the European Union's Horizon 2020 research and innovation programme under grant agreement No. 646804-ERC-COG-BNYQ. K.V.M. acknowledges partial support via Andrew and Erna Finci Viterbi Fellowship.}
}

\maketitle

\begin{abstract}
We present the design and hardware implementation of a radar prototype that demonstrates the principle of a sub-Nyquist collocated multiple-input multiple-output (MIMO) radar. The setup allows sampling in both spatial and spectral domains at rates much lower than dictated by the Nyquist sampling theorem. Our prototype realizes an X-band MIMO radar that can be configured to have a maximum of 8 transmit and 10 receive antenna elements. We use frequency division multiplexing (FDM) to achieve the orthogonality of MIMO waveforms and apply the Xampling framework for signal recovery. The prototype also implements a cognitive transmission scheme where each transmit waveform is restricted to those pre-determined subbands of the full signal bandwidth that the receiver samples and processes. Real-time experiments show reasonable recovery performance while operating as a $4\times5$ thinned random array wherein the combined spatial and spectral sampling factor reduction is $87.5$\% of that of a filled $8\times10$ array. 
\end{abstract}

\begin{IEEEkeywords}
MIMO radar, sub-Nyquist, compressed sensing, collocated, cognitive radar
\end{IEEEkeywords}

\section{Introduction}
\label{sec:intro}
Multiple input multiple output (MIMO) radar is a novel radar paradigm \cite{fishler2004mimo} that uses an array of several transmit and receive antenna elements, with each transmitter radiating a different waveform. In a \textit{collocated} MIMO radar \cite{li2007mimo}, the antenna elements are placed close to each other so that the radar cross-section of a target appears identical to all the elements. The waveform diversity in a collocated MIMO is based on the mutual orthogonality - usually in time, frequency or code - of different transmitted signals. The receiver separates and coherently processes the target echoes corresponding to each transmitter. The angular resolution of MIMO is same as a \textit{virtual} phased array with the same antenna aperture but many more antenna elements than MIMO.

While a radar achieves high angular resolution by using a large virtual aperture, its range-time resolution can be improved by transmitting signals with large bandwidth. In other words, the conventional processing resolution is limited by the number of elements and the receiver sampling rate. Several methods have been proposed to address the problem of preserving the MIMO radar resolution when either the number of antennas \cite{rossi2014spatial} or the number of received samples \cite{yu2010mimo, kalogerias2014matrix, mimoMC} is reduced. Most exploit the fact that the target scene is \textit{sparse} facilitating the use of compressed sensing (CS) methods \cite{CSBook,SamplingBook}.

Recently, \cite{cohen_MIMO2016} proposed a \textit{sub-Nyquist collocated MIMO radar} (sub-Nyquist MIMO hereafter) that can recover the target range and azimuth by simultaneously \textit{thinning} an antenna array and sampling received signals at sub-Nyquist rates. The recovery algorithm uses the \textit{Xampling} framework where Fourier coefficients of the received signal are acquired from their low-rate samples (or \textit{Xamples}) \cite[p. 387-388]{SamplingBook}\cite{bar2014sub}. Application of Xampling in space and time enables sub-Nyquist sampling without loss of any of the aforementioned radar resolutions. The Xamples are expressed as a matrix of unknown target parameters and the reconstruction algorithm is derived by extending the orthogonal matching pursuit (OMP) \cite{SamplingBook} to simultaneously solve a system of CS matrix equations. In sub-Nyquist MIMO, the radar antenna elements are randomly placed within the aperture (see \cite{lo1964random} for introduction and \cite{rossi2014spatial} for recent updates on random arrays), and signal orthogonality is achieved by frequency division multiplexing (FDM). In a conventional MIMO radar, the use of non-overlapping FDM waveforms results in a strong range-azimuth coupling \cite{rabaste2013signal, sun2014analysis,xu2015joint} in the receiver processing, and therefore, it is common to use orthogonal code signals (i.e. code division multiplexing or CDM). The coupling due to FDM can be reduced by randomizing the carrier frequencies across transmitters \cite{sun2014analysis}. The FDM-based sub-Nyquist MIMO mitigates the range-azimuth coupling by randomizing the element locations in the aperture.

In this work, we present the first hardware prototype of a sub-Nyquist MIMO that can demonstrate reduction in both spatial and spectral sampling using real-time analog signals. This implementation follows the recommendations of \cite{cohen_MIMO2016} for signal orthogonality, array structure and reconstruction algorithms. The prototype can be configured either as a filled or thinned array, thereby allowing comparison of Nyquist and sub-Nyquist spatial sampling using the same hardware. Our previous work in \cite{radar_demo} presented the hardware realization of spectral sub-Nyquist sampling in radar. In \cite{radar_demo}, a few randomly chosen, narrow subbands of the received signal spectrum are pre-filtered before being sampled by low-rate analog-to-digital converters (ADCs). Since this implementation uses a bandpass filter and an ADC for each subband, a similar implementation of spectral sub-Nyquist sampling in each channel of a MIMO receiver would require enormous hardware resources. To circumvent such a simplistic and excessively large design, we instead transmit only in those subbands that are sampled by the receiver. This eliminates all pre-filtering band-pass stages and each receiver channel requires only one low-rate ADC to sample all subbands. 

Limiting the signal transmission to selective subbands allows for more in-band power resulting in an increase in signal-to-noise ratio (SNR). This approach has recently been proposed in the context of sub-Nyquist \textit{cognitive} radars \cite{cohen_cognitive2016}. Our prototype, therefore, additionally demonstrates application of cognitive transmission in sub-Nyquist MIMO radar. Additionally, a radar that transmits in several disjoint narrow bands has advantages of disguising the transmit frequencies as an effective electronic counter measure (ECM), and also escape radio-frequency (RF) interference from other licensed radiators in the vacant non-transmit subbands.

In the following section, we briefly review the signal and system model of sub-Nyquist MIMO. We then describe the design philosophy of our prototype and its major sub-modules in Section III. Finally, we present results obtained by the prototype in real-time experiments.

\section{Sub-Nyquist Collocated MIMO Radar}
With the exception of cognitive transmission, the array and signal models of sub-Nyquist MIMO realized by our prototype closely follow that detailed by \cite{cohen_MIMO2016} and, hence, we only summarize them here.
\label{sec:mimo}
\subsection{MIMO Radar Model}
\label{subsec:model}
Let the operating wavelength of the radar be $\lambda$ and the total number of transmit and receive elements be $T$ and $R$ respectively. The classic approach to collocated MIMO adopts a virtual uniform linear array (ULA) structure \cite{chen2009signal}, where the receive antennas spaced by $\frac{\lambda}{2}$ and transmit antennas spaced by $R\frac{\lambda }{2}$ form two ULAs (or vice versa, see e.g. \cite{brookner2015mimo}). Here, the coherent processing of a total of $TR$ channels in the receiver creates a virtual equivalent of a phased array with $TR$ $\frac{\lambda }{2}$-spaced receivers and normalized aperture $Z=\frac{TR}{2}$.

Let us now consider a collocated MIMO radar system that has $M<T$ transmit and $Q<R$ receive antennas. The locations of these antennas are chosen uniformly at random within the aperture of the virtual array mentioned above. The $m$th transmitting antenna sends a unique pulse ${{s}_{m}}\left( t \right)$ given by\par\noindent\small
\begin{equation}
\label{eq:trMth}
    \begin{array}{lll}
     {{s}_{m}}\left( t \right)={{{h}_{m}}\left( t\right)}{{e}^{j2\pi {{f}_{c}}t}},\quad0\le t\le \tau,
    \end{array}
\end{equation}\normalsize
    where $\tau$ denotes the pulse repetition interval (PRI), ${{f}_{c}}$ is the common carrier frequency at the RF (radio-frequency) stage, and $\{{{h}_{m}}\left( t \right)\}_{m=0}^{M-1}$ is a set of narrowband, orthogonal FDM pulses each with the continuous-time Fourier transform (CTFT)\par\noindent\small
\begin{equation}
\label{eq:ctft}
H_m(\omega) = \int\limits_{-\infty}^{\infty}h_m(t)e^{-j\omega t}dt.
\end{equation}\normalsize
Suppose the target scene consists of $L$ non-fluctuating point targets (Swerling-0 model) \cite{skolnik} whose location is given by their range-time $\tau_l$ and direction-of-arrival (DoA) $\theta_l$, $1 \le l \le L$. The pulses transmitted by the radar are reflected back by the targets and collected at the receive antennas. When the received waveform is downconverted from RF to baseband, we obtain the following signal at the $q$th antenna,\par\noindent\small
\begin{equation}
x_q \left( t \right) = \sum\limits_{m=0}^{M-1} \sum\limits_{l=1}^{L} \alpha_l h_m \left( t-\tau _{l} \right) e^{j2 \pi \beta_{mq} \theta _l},
\end{equation}\normalsize
where $\alpha_l$ denotes the complex-valued reflectivity of the $l$th target and $\beta_{mq}$ is governed by the array structure. Our goal is to estimate the range-time ${{\tau}_{l}}$ and azimuths ${{\theta }_{l}}$ of each target.
\subsection{Sub-Nyquist Range-Azimuth Recovery}
The application of Xampling in both space and time enables recovery of range and direction at sub-Nyquist rates. The performance guarantees of this procedure are provided in \cite{cohen_MIMO2016}. The received signal $x_q(t)$ is separated into $M$ channels, aligned and normalized. The Fourier coefficients of the received signal corresponding to the channel that processes the $m$th transmitter echo are given by\par\noindent\small
\begin{equation}
\label{coffAlligned}
y_{m,q}[k]=   \sum_{l=1}^{L} \alpha_l e^{j2\pi \beta_{mq} \theta_l}e^{-j\frac{2\pi }{\tau}k\tau_l} e^{-j2\pi f_m \tau_l},
\end{equation}\normalsize
where $-\frac{N}{2} \leq k \leq -\frac{N}{2}-1$, $f_m$ is the (baseband) carrier frequency of the $m$th transmitter and $N$ is the number of Fourier coefficients per channel. Xampling obtains a set $\kappa$ of arbitrarily chosen Fourier coefficients from low rate samples of the received channel signal such that $|\kappa| = K < N$.

Let us now consider the $m$th transmission. Suppose $\bm{Y}^m$ is the $K \times Q$ matrix with $q$th column given by $y_{m,q}[k]$, $k \in \kappa$. The matrix $\bm{Y}^m$ can be expressed as\par\noindent\small
\begin{equation}
\label{eq:model}
\bm{Y}^m = \bm{A}^m \bm{X} \left(\bm{B}^m\right)^T,
\end{equation}\normalsize
where $\bm{X}$ is a sparse matrix in which the location and values of the non-zero elements correspond to the locations and reflectivity of the targets respectively. The matrices $\bm{A}^m$ and $\bm{B}^m$ are known functions of radar parameters ($T$, $R$, $f_m$, $\kappa$, $\tau$, and transmit bandwidth), and each of their columns correspond to a range and an azimuth cell, respectively. The sparse matrix $\bm X$ can be recovered from the set of equations (\ref{eq:model}) for all $0 \leq m \leq M-1$, by solving the optimization problem\par\noindent\small
\begin{align}
\label{eq:opt}
 & \underset{\bm{X}}{\text{minimize}}\;\; ||\bm{X}||_0 \nonumber\\
 & \text{subject to} \;\; \bm{Y}^m = \bm{A}^m \bm{X} \left( \bm{B}^m\right)^T, \quad 0 \leq m \leq M-1.
\end{align}\normalsize
An approximate solution to this problem can be obtained through an extension of the matrix OMP method \cite{wimalajeewa2013recovery}. We refer the reader to \cite{cohen_MIMO2016} for full details of this recovery algorithm. 
\subsection{Cognitive Transmission}
\label{subsec:cognitive}
\begin{figure}
  \includegraphics[scale=0.25]{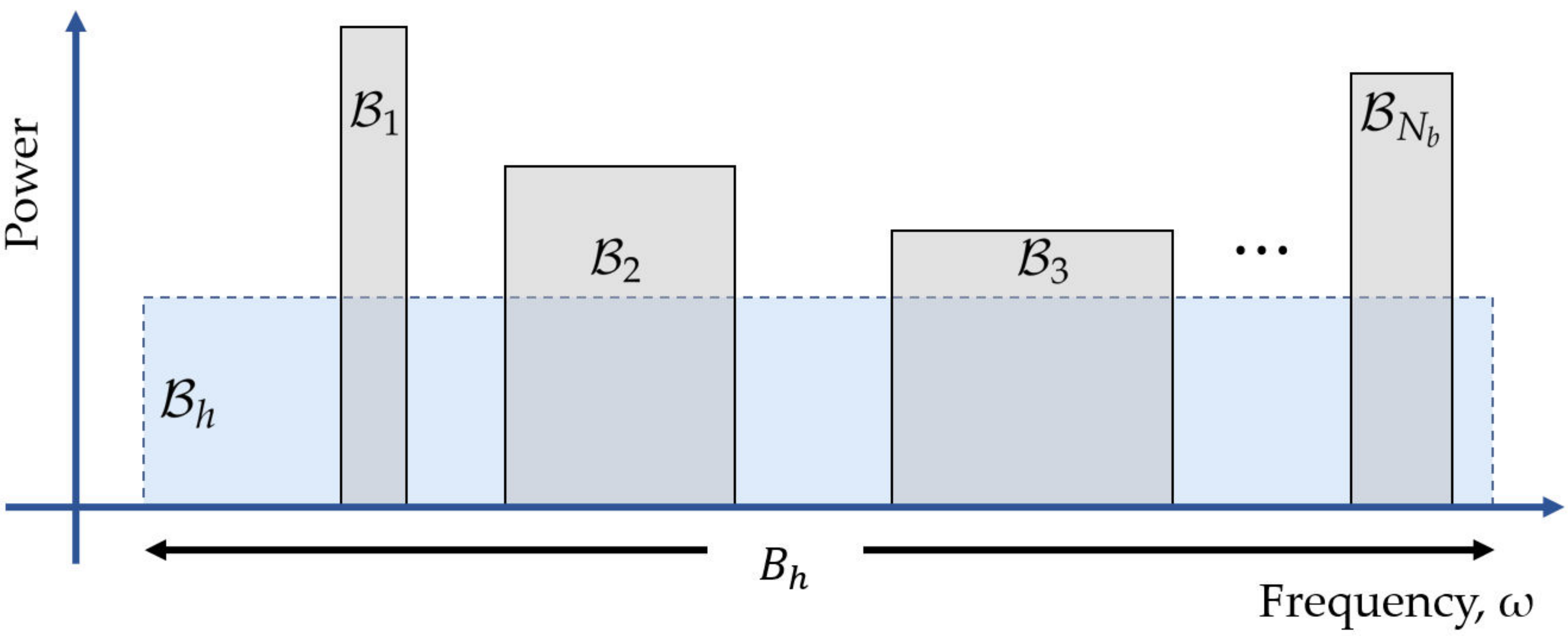}
  \caption{\hspace{-4pt}A conventional radar with signal bandwidth $B_h$ transmits in the  band $\mathcal{B}_h$. A cognitive radar transmits only in subbands $\{\mathcal{B}_i\}_{i=1}^{N_b}$, but with more in-band power than the conventional radar.\vspace{-9pt}}
	\label{fig:cogspec}
\end{figure}
Let $\mathcal{B}_h$ be the set of all frequencies in the transmit signal spectrum of effective bandwidth $B_h$. In the cognitive radar transmission, the spectrum $\tilde{H}_m(\omega)$ of each of the transmitted waveforms $\tilde{h}_m(t)$ is limited to a total of $N_b$ non-overlapping frequency bands $\mathcal{B}_i$, $1 \le i \le N_b$ (Figure~\ref{fig:cogspec}):\par\noindent\small
\begin{align}
\tilde{H}_m(\omega)
&= \begin{dcases} 
    \gamma(\omega) H_m(\omega), \phantom{1}\phantom{1} \omega \in \bigcup_{i=1}^{N_b}\mathcal{B}_i \subset \mathcal{B}_h\\
    0, \phantom{1}\phantom{1}\text{otherwise}
   \end{dcases}
\end{align}\normalsize
where $\gamma(\omega)={B_h}/{|\mathcal{B}_i}|$ for $\omega \in \mathcal{B}_i$. The total transmit power $P_t$ remains the same such that the power relation between the conventional and cognitive waveforms is\par\noindent\small
\begin{align}
\int_{-B_h/2}^{B_h/2} |H_m(\omega)|^2\, \mathrm{d}\omega = \sum_{i=1}^{N_b}\int\limits_{\mathcal{B}_i} |\tilde{H}_m(\omega)|^2\, \mathrm{d}\omega = P_t
\end{align}\normalsize
In a cognitive radar, the sub-Nyquist receiver obtains the set $\kappa$ of the Fourier coefficients only from the subbands $\mathcal{B}_i$.
\section{Hardware Design}
\begin{table}
\centering
\caption{\hspace{-10pt}Technical characteristics of the prototype}
\label{tbl:techmodes}
\vspace{13pt}
	\begin{tabular}{ l | c | c | c | c}
		\hline
         \noalign{\vskip 1pt}    
         	Parameters & Mode 1 & Mode 2 & Mode 3 & Mode 4\\[1pt]
		\hline
		\hline
        \noalign{\vskip 1pt}    
		   	\#Tx, \#Rx & 8,10 & 8,10 & 4,5 & 8,10\\[1pt]
		   	Element placement & Uniform & Random & Random & Random\\[1pt]
		   	Equivalent aperture & 8x10 & 8x10 & 8x10 & 20x20\\[1pt]            
            Angular resolution (sine of DoA) & 0.025 & 0.025 & 0.025 & 0.005\\[1pt]
		\hline
        \noalign{\vskip 1pt}    		            
	        Range resolution & \multicolumn{4}{c}{1.25 m}\\[1pt]
            Signal bandwidth per Tx & \multicolumn{4}{c}{12 MHz (15 MHz including guard-bands)}\\[1pt]
		   	Pulse width & \multicolumn{4}{c}{4.2 $\mu$s}\\[1pt]
            Carrier frequency & \multicolumn{4}{c}{10 GHz}\\[1pt]
            Unambiguous range & \multicolumn{4}{c}{15 km}\\[1pt]
            Unambiguous DoA & \multicolumn{4}{c}{180$^{\circ}$ (from -90$^{\circ}$ to 90$^{\circ}$)} \\[1pt]
		\hline
		\hline
	\end{tabular}
\end{table}
\begin{figure}[!t]
\centering
  \includegraphics[scale=0.3]{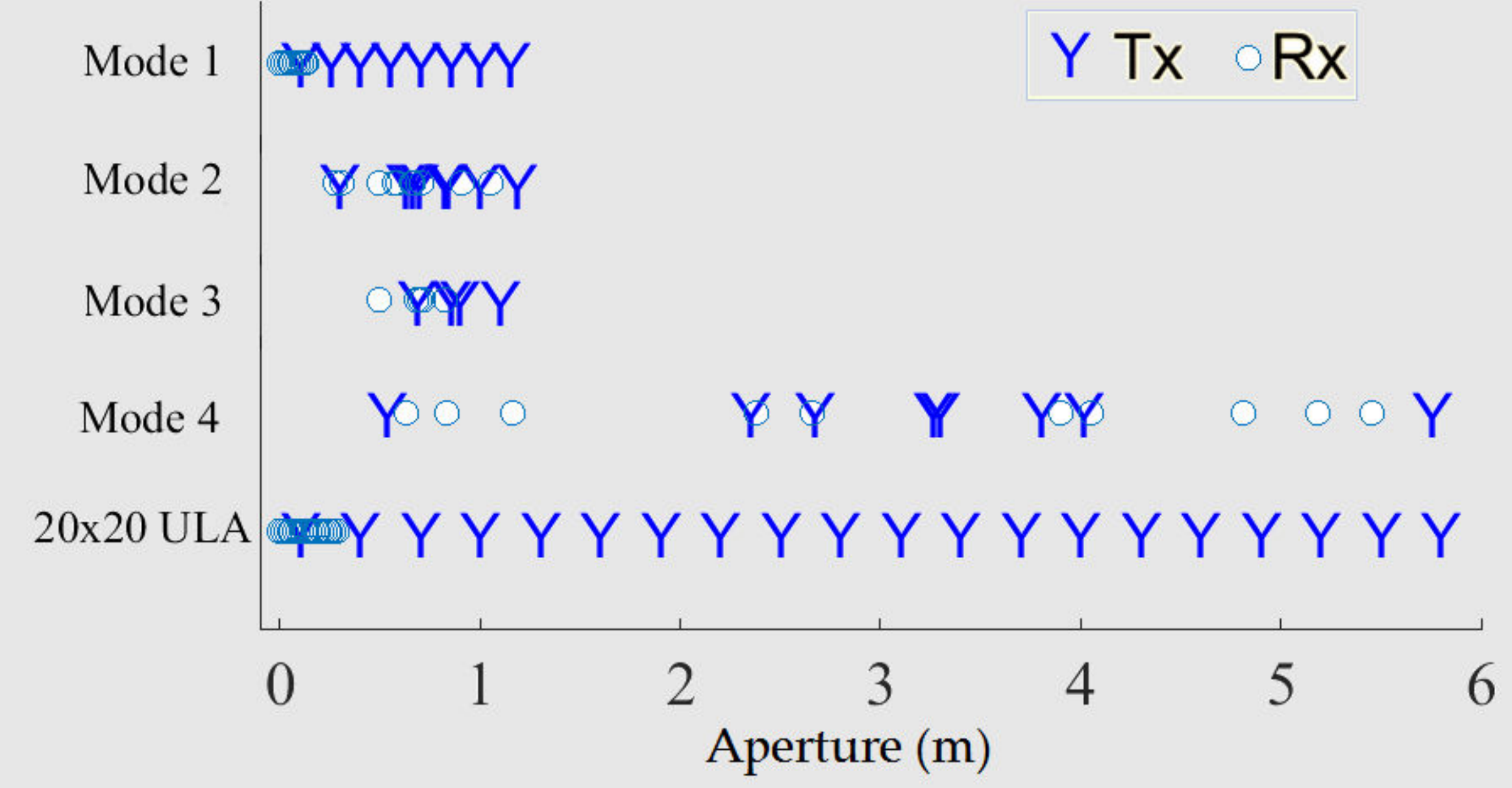}
  \caption{\hspace{-4pt}Tx and Rx element locations for the hardware prototype modes over a 6 m antenna aperture. Mode 4's virtual array equivalent is the $20\times20$ ULA.\vspace{-9pt}}
\label{fig:arrays}
\end{figure}
We consider the implementation of a MIMO radar architecture that has a maximum of 8 transmit (Tx) and 10 receive (Rx) antenna elements. As explained in subsequent sections, the spatial and spectral sampling aspects of sub-Nyquist MIMO that we intend to demonstrate manifest only in the receiver processing. Therefore, our radar \textit{prototype} does not physically radiate the transmit waveforms from an antenna and receive data from actual targets in real-time. Instead, multiple transmit waveforms are pre-computed externally at baseband, their echoes from simulated targets are recorded and the complex samples (in-phase $I$ and quadrature-phase $Q$ pairs) are stored in an on-board memory of a custom-designed waveform generator board. The prototype then processes these pre-recorded signals in real-time. Similarly, we omit the implementation of the up-conversion to RF carrier frequency in the transmitter and the corresponding down-conversion in the receiver from this prototype. We would assume that the physical array aperture and simulated target response correspond to an X-band ($f_c = 10$ GHz) radar. The choice of radar frequency band also affects the clutter response that we intend to consider in a future extension of this prototype.
\subsection{Design Philosophy}
\label{subsec:designphil}
A conventional 8x10 MIMO radar receiver would require simultaneous hardware processing of 80 (or 160 I/Q) data streams. Since a separate sub-Nyquist receiver for each of these 80 channels is expensive, we implement the eight channel analog processing chain for only one receive antenna element and serialize the received signals of all 10 elements through this chain. This approach allows the prototype to implement a number of receivers greater than 10 as the eight-channel hardware only limits the number of transmitters.

Given a particular receive element, we intend to extract Fourier coefficient set $\kappa$ for each of its transmit channels using low-rate ADCs. It has been shown \cite[pp. 210-268]{CSBook} that high recovery performance is promised when these coefficients are drawn uniformly at random. An ADC can not, however, individually acquire each of the randomly chosen Fourier coefficients. Therefore, sub-Nyquist radar prototype in \cite{radar_demo} opted for sampling random disjoint subsets of $\kappa$, with each subset containing consecutive Fourier coefficients. The prototype in \cite{radar_demo} used four random Fourier coefficient groups, pre-filtered the baseband signal to corresponding four subbands (or Xampling \textit{slices}), and sampled each subband via separate low-rate ADC.

If we use the same pre-filtering approach as in \cite{radar_demo} for each of the eight channels of our sub-Nyquist MIMO prototype, then the hardware design would need a total of $4\times8=32$ bandpass filters (BPFs) and ADCs excluding the analog filters to separate transmit channels. We sidestep this requirement by adopting cognitive transmission wherein the analog signal of each channel lives only in certain pre-determined subbands and consequently, a BPF stage is not required. More importantly, for each channel, a single low-rate ADC can \textit{subsample} this narrow-band signal as long as the subbands are \textit{coset} bands so that they don't aliase after sampling \cite{cohen2014channel}. This implementation needs only eight low-rate ADCs, one per channel. Another advantage of this approach is flexibility of the prototype in selecting the Xampling \textit{slices}. Unlike \cite{radar_demo}, the number and spectral locations of slices are not permanently fixed, and they can be changed (within the constraints of aliasing due to subsampling).

The prototype can be configured to operate in various array configurations or \textit{modes}. When operating at its maximum strength of 8 Tx and 10 Rx elements, it can be programmed as either a ULA (Mode 1) or a random array (Mode 2), each with the equivalent aperture of an $8\times10$ virtual array, i.e., $1.2$ m. For the same aperture, the system can be operated as a thinned $4\times5$ array (Mode 3). In this configuration fewer receivers are serialized and the channels corresponding to the removed transmit elements are not processed by the digital receiver. Mode 3, hence, demonstrates the spatial sub-Nyquist sampling. Finally, the prototype can also function as a $8x10$ thinned array (Mode 4) which can be viewed as a spatial sub-Nyquist version of a $20x20$ virtual array with aperture of $6$ m.

Figure~\ref{fig:arrays} shows exact details of element locations for all four modes. Table \ref{tbl:techmodes} summarizes the technical parameters of the prototype for all four array configurations. As mentioned before, this system employs FDM-based signal orthogonality with each transmit signal $h_m(t)$ chosen to be approximately flat in spectrum, over the extent of 12 MHz (one-sided band). Each of the transmit waveforms is separated from its nearest neighbor by a 3 MHz guard-band.

\subsection{System Description}
\begin{figure*}[!t]
\centering
\includegraphics[width=\textwidth]{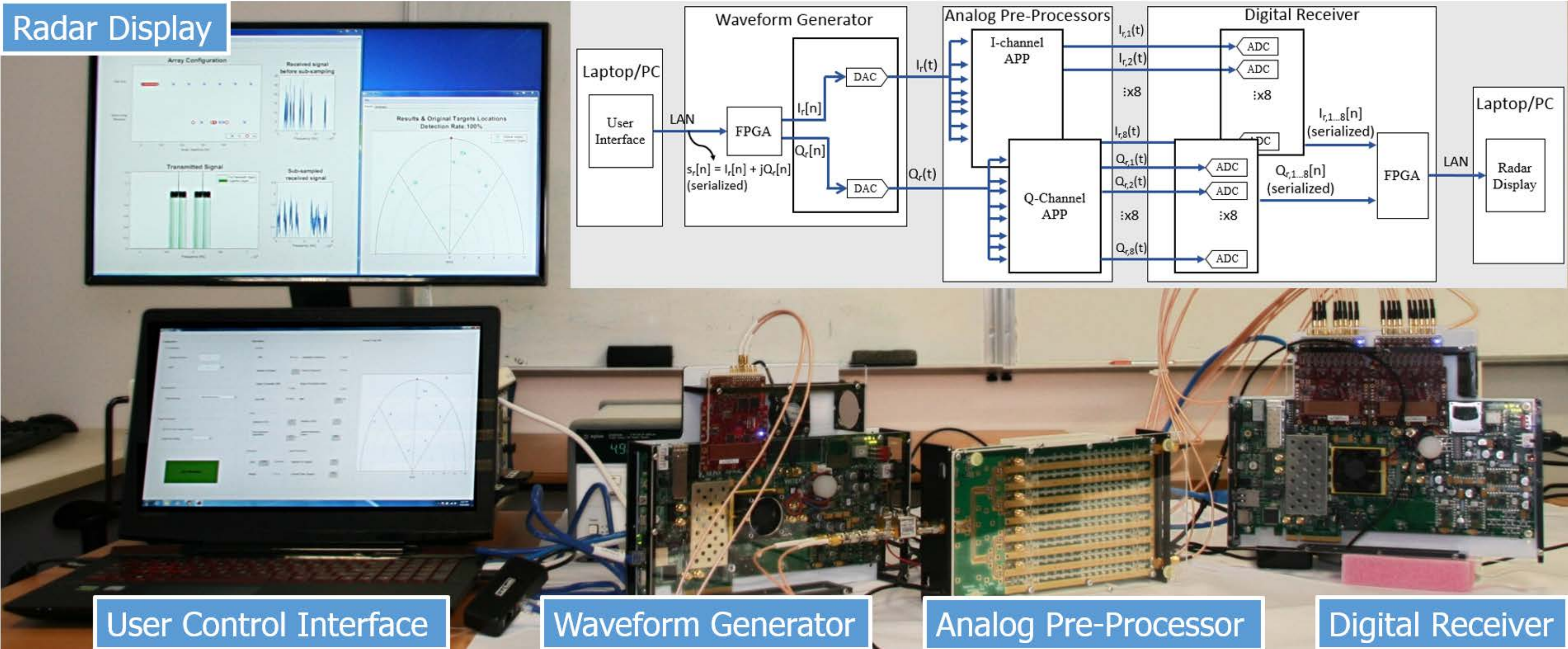}
\caption{\hspace{-4pt}Sub-Nyquist MIMO prototype and user interface. The analog pre-processor module consists of two APP cards mounted on opposite sides of a common chassis. The inset shows the simplified block diagram of the system. The subscript $r$ represents received signal samples for $r$th receiver. Wherever applicable, the second subscript corresponds to a particular transmitter. The square brackets (parentheses) are used for digital (analog) signals.\vspace{-10pt}}
\label{fig:mimoblockphoto}
\end{figure*}
Figure~\ref{fig:mimoblockphoto} shows the sub-Nyquist MIMO prototype, user interface and radar display. The inset graph depicts the signal flow through a simplified block diagram. The user selects the prototype mode from the control interface and passes the control triggers to the \textit{transmit waveform generator} card, where an FPGA device reads out the pre-stored received waveform from an on-board memory. Two separate digital-to-analog converters (DACs) - one each for $I$ and $Q$ samples - convert the resulting signal to baseband analog domain. The transmit waveform generator is off-the-shelf Xilinx VC707 evaluation board that is custom fit with a 4DSP FMC204 16-bit DAC card. Each of the $I$ and $Q$ analog signals are then passed on to their respective analog pre-processor cards.

A custom-built analog pre-processor (APP) splits the 120 MHz baseband analog signal from the waveform generator in 8 channels. The 9 dB attenuation due to 8-channel splitter is compensated with the use of 10 dB amplifier for each channel. The signal corresponding to each transmitter is then filtered using BPFs with 12 MHz passband. Only the first transmitter channel uses a low-pass filter as it is difficult to practically realize a bandpass filter with a passband close to zero. The first five channels use Chebyshev filter design and the rest are elliptic filters, all with a passband ripple of 0.1 dB. Since subsampling raises the out-of-band noise, all of these front-end filters are designed to provide approximately 30 dB stopband attenuation. The imbalance in gain and spectral distortion are corrected by placing tunable equalizers at the end of APP chain. The channelized $I/Q$ analog signals are then digitized using low-rate 16-bit ADCs in a digital receiver card. A digital receiver is realized using a single Xilinx VC707 evaluation board with two eight-channel 4DSP FMC168 digitizer daughter cards, one each for $I$ and $Q$ signals. The digital receiver output is transferred over LAN to a radar display. 
\begin{figure}[!t]
\centering
\subfloat[Before subsampling]{%
  \includegraphics[width=4.1cm]{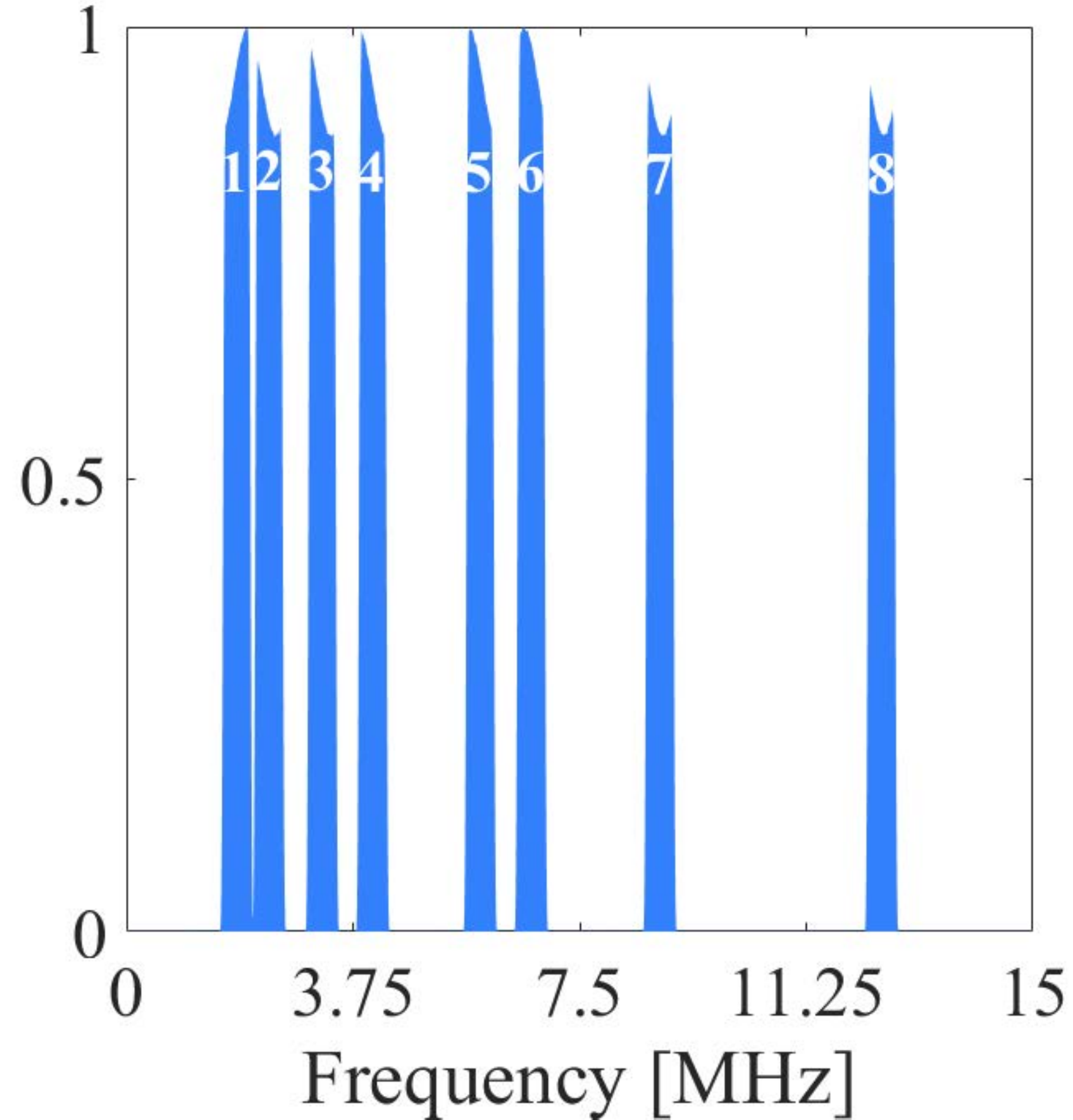}%
  \label{fig:beforesub}%
}\qquad
\subfloat[After subsampling]{%
  \includegraphics[width=4.1cm]{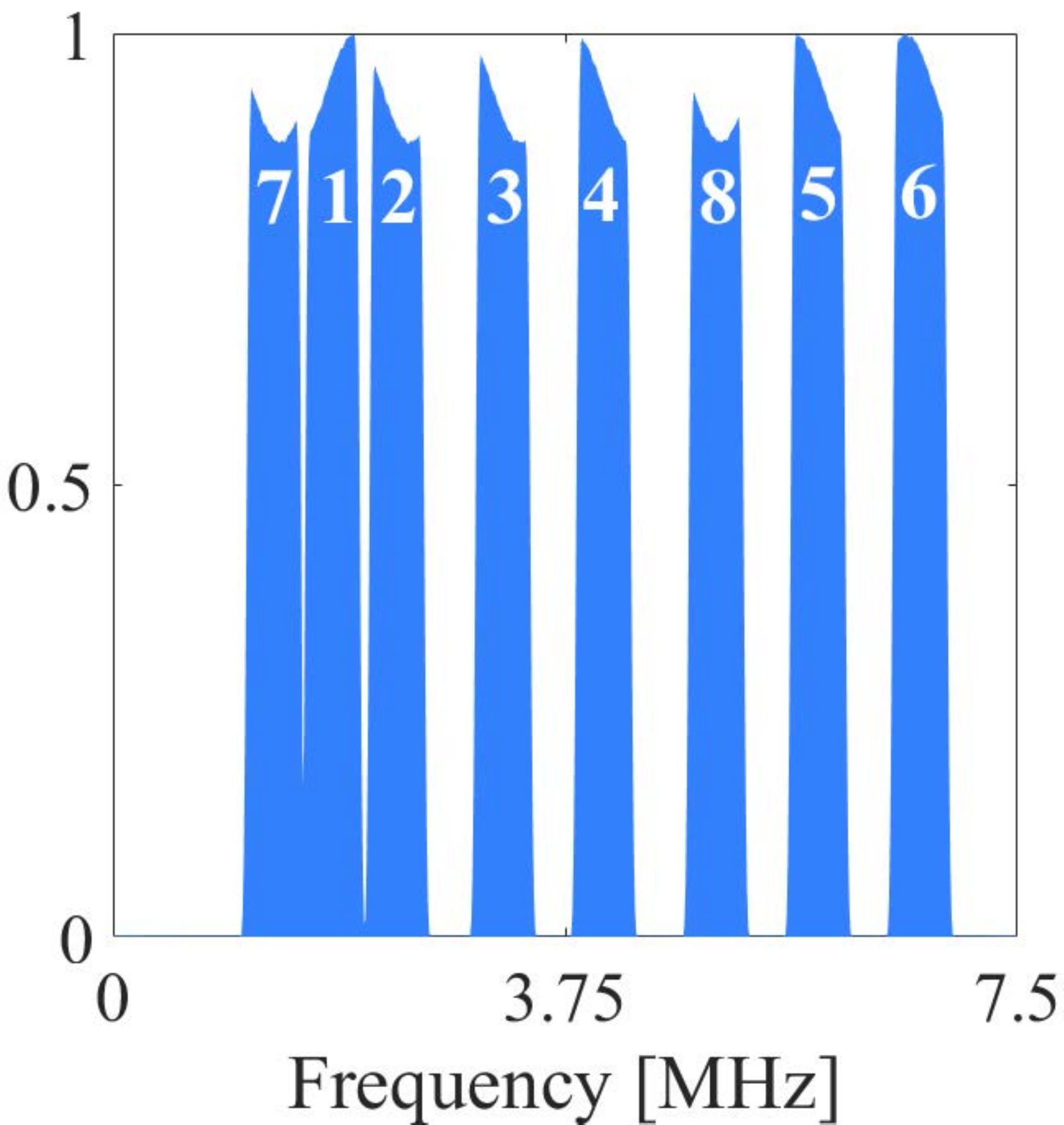}%
  \label{fig:aftersub}%
}
\caption{\hspace{-5pt}The normalized one-sided spectrum of one channel of a given receiver (a) before and (b) after subsampling with a 7.5 MHz ADC. Each of the subbands span 375 kHz and is marked with a numeric label. In a non-cognitive processing, the signal occupies the entire 15 MHz spectrum before sampling.\vspace{-14pt}}
\label{fig:slicespec}
\end{figure}
\begin{figure}[!t]
\centering
\includegraphics[width=\columnwidth]{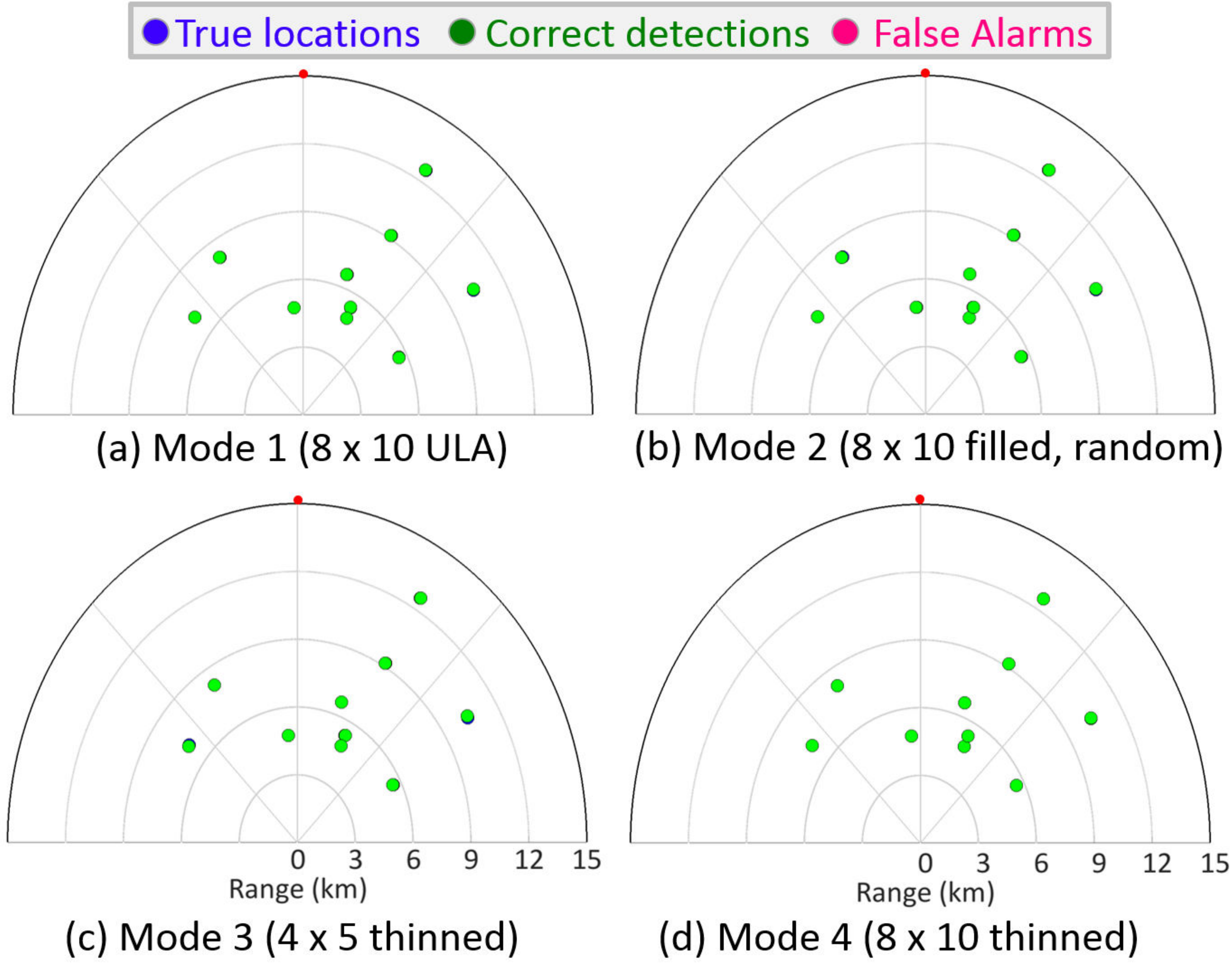}
\caption{\hspace{-6pt}Plan Position Indicator (PPI) display of results for Mode 1 and 3. The origin is the location of the radar. The red dot indicates the north direction relative to the radar. Positive (negative) distances along the horizontal axis correspond to the east (west) of the radar. Similarly, positive (negative) distances along the vertical axis correspond to the north (south) of the radar. The estimated targets are plotted over the ground truth. \vspace{-14pt}}
\label{fig:sameperf}
\end{figure}
As shown in Figure~\ref{fig:beforesub}, the cognitive radar signal occupies only certain subbands in a 15 MHz band. Here, the \textit{sliced} transmit signal has eight subbands each of width 375 kHz with the frequency range of 1.63-2, 2.16-2.53, 3.05-3.42, 3.88-4.25, 5.66-6.03, 6.51-6.88, 8.64-9.01 and 12.32-12.69 MHz before subsampling. The resulting coherence \cite{xia2005achieving} for this selection of Fourier coefficients is 0.42. The total signal bandwidth is $0.375 \times 8 = 3$ MHz. This signal is subsampled at 7.5 MHz and, as shown in Figure~\ref{fig:aftersub}, there is no aliasing between different subbands. A non-cognitive signal would have occupied entire 15 MHz spectrum requiring a Nyquist sampling rate of 30 MHz. Therefore, use of cognitive transmission enables spectral sampling reduction by a factor of $4$ ($=30$ MHz$/7.5$ MHz) for each channel. Depending on whether the guard-bands of non-cognitive transmission are included in the computation or not, the effective signal bandwidth is reduced by a factor of $5$ ($=15$ MHz$/3$ MHz) or $4$ ($=12$ MHz$/3$ MHz) respectively for each channel. Mode 3 has 50\% spatial sampling reduction when compared with Mode 1 or 2. If we account for both spatial and spectral sampling reduction in Mode 3, then we use a total one-eighth of the Nyquist sampling rate and one-tenth of the Nyquist signal bandwidth (guard-bands included). The sampling rate reduction is, therefore, seven-eighth or $87.5\%$ in Mode 3. The receiver processes $80$ and $20$ channels in $8x10$ and $4x5$ arrays, respectively. So, the hardware resources are reduced by $75\%$ in Mode 3.
\subsection{Experimental Results}
\label{subsec:results}
We evaluated the performance of all the modes through hardware simulations. Only one pulse was transmitted in all experiments and all modes were evaluated against the identical target scenarios. We injected the received signal corresponding to the echoes from 10 targets, placed at arbitrary range and azimuths, in the transmit waveform generator. In the first experiment, when the \textit{minimum} angular spacing (in terms of the sine of DoA) between any two targets was approximately $0.025$, the recovery performance of the thinned $4\times5$ array in Mode 3 was not worse than Modes 1 and 2 until the noise was dramatically increased. Figure~\ref{fig:sameperf} shows the detection performance of all the modes for this experiment when the signal SNR = $-5$ dB; the injected noise is complex white Gaussian. Here, a successful detection (green circle) occurs when the estimated target is within two range cells and one DoA bin of the ground truth (blue circle); otherwise, the estimated target is labeled as a false alarm (magenta circle). In case of high SNR or absence of noise, our criterion for successful detection is \textit{sensu stricto}, i.e. the estimated target must lie at the exact location of the ground truth for a successful detection.

\begin{figure}[!t]
\centering
\includegraphics[width=\columnwidth]{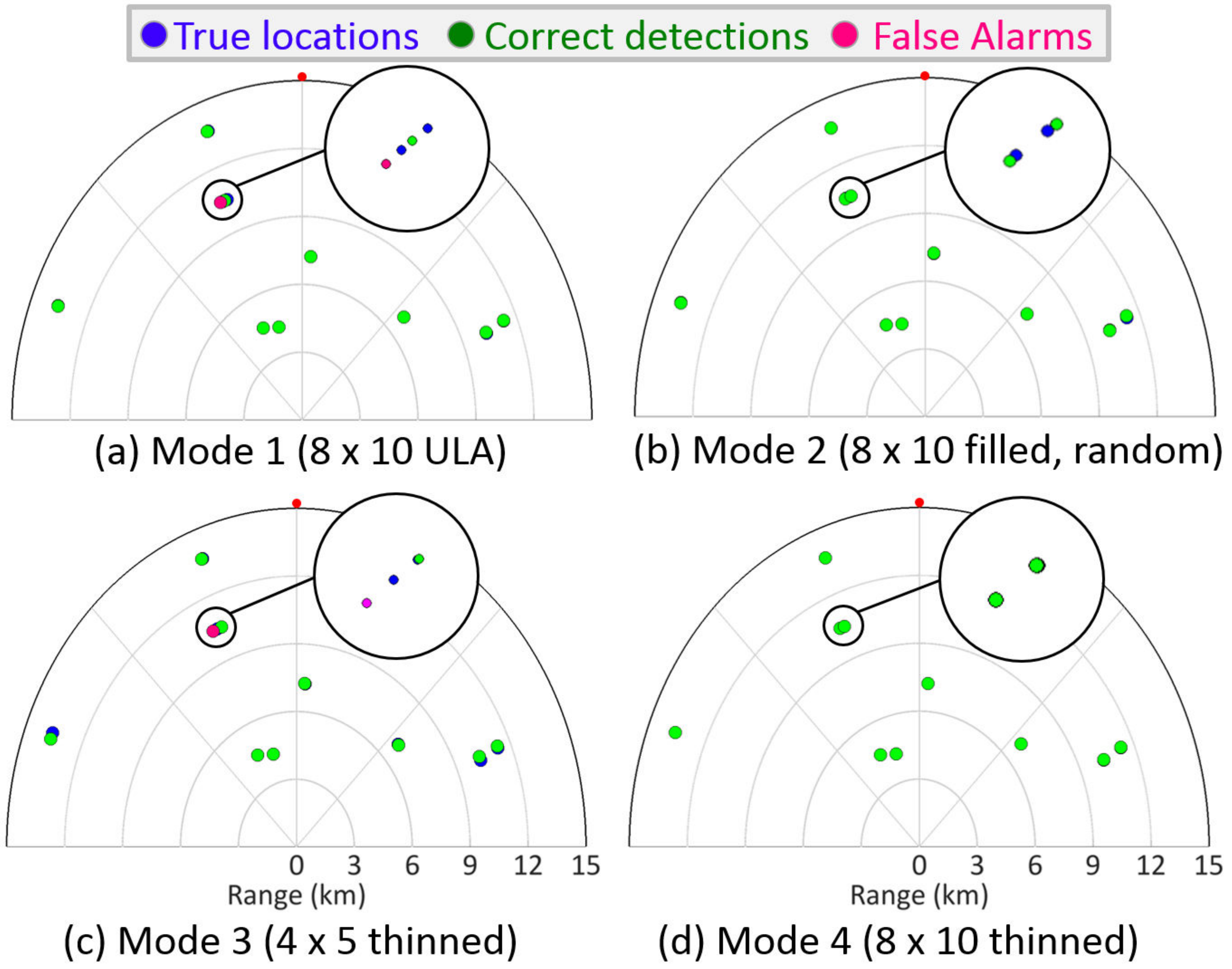}
\caption{\hspace{-6pt}As in Fig. 5, but for a closely-spaced target scenario. The inset plots show the selected region in each PPI display at a magnified scale.\vspace{-9pt}}
\label{fig:diffperf}
\end{figure}

In the second experiment, the minimum angular spacing between the two targets was reduced to $0.02$, and the SNR of the injected signal remained at $-5$ dB. Since the angular resolution of Mode 4 is better than the other three modes (see Table~\ref{tbl:techmodes}), all the targets are detected successfully in Mode 4. Mode 1 and 3 showed a false alarm as seen in the inset plots of Figure~\ref{fig:diffperf}. The Mode 2 also shows successful recovery in the broad sense of our detection criterion. The strict sense location error in Mode 2 is clearly larger than that in Mode 4. However, relatively better performance of Mode 2 over Modes 1 and 3 is not entirely fortuitous here. Figure~\ref{fig:arrays} shows that both Tx and Rx array elements in Mode 2 are distributed such that its virtual array is wider than Modes 1 and 3. Thus, the effective angular resolution for Mode 2 could be better than 1 and 3, but still worse than 4.
\section{Summary}
\label{sec:summary}
We presented the first hardware prototype of sub-Nyquist MIMO that demonstrates real-time operation of both spatial and spectral reduction in sampling. The thinned 4x5 array achieves the detection performance of its filled array counterparts even though the combined reduction of spatial and spectral sampling is $87.5\%$. While we did not analyze the performance for Doppler recovery and clutter contamination, the prototype design does not restrict such an evaluation. Future theoretical insights on the selection of best subbands and improved recovery algorithms can further enhance the performance.
\balance
\bibliographystyle{IEEEtran}
\bibliography{refs}

\end{document}